\documentclass[useAMS,usenatbib,usegraphicx]{pasj00}

\RelativeFigurePath{figs-lowres}

\begin{document}

\SetRunningHead{A.\,T. Okazaki et al.}
               {Hydrodynamic Interaction in the TeV Binary PSR~B1259$-$63/LS~2883}

\title{Hydrodynamic Interaction between 
the Be Star and the Pulsar in the TeV Binary PSR B1259$-$63/LS~2883}


\author{Atsuo T. \textsc{Okazaki}} 
\affil{Faculty of Engineering, Hokkai-Gakuen University, Toyohira-ku,
       Sapporo 062-8605, Japan}
\email{okazaki@elsa.hokkai-s-u.ac.jp}
\author{Shigehiro \textsc{Nagataki}} 
\affil{Yukawa Institute for Theoretical Physics, Oiwake-cho,
       Kitashirakawa, Sakyo-ku, Kyoto 606-8502, Japan}
\author{Tsuguya \textsc{Naito}} 
\affil{Faculty of Management Information, Yamanashi Gakuin University,
       Kofu, Yamanashi 400-8575, Japan}
\author{Akiko \textsc{Kawachi}} 
\affil{Department of Physics, Tokai University, Hiratsuka, Kanagawa
       259-1292, Japan}
\author{Kimitake \textsc{Hayasaki}}
\affil{Department Astronomy, Kyoto University, Oiwake-cho, Kitashirakawa, 
       Sakyo-ku, Kyoto 606-8502, Japan}
\author{Stanley P. \textsc{Owocki}} 
\affil{Bartol Research Institute, University of Delaware, 
       Newark, DE 19716, USA}
       \and
\author{Jumpei \textsc{Takata}} 
\affil{Department of Physics, University of Hong Kong, Pokfulam Road, Hong Kong, 
       China}



\KeyWords{
gamma rays: theory -- stars: emission-line, Be -- stars: winds, outflows --
stars: individual (B~1259$-$63)
}

\maketitle


\begin{abstract}
We study the interaction between the Be star and the pulsar in the TeV binary PSR~B1259$-$63/LS~2883, using 3-D SPH simulations of the tidal and wind interactions in this Be-pulsar system. We first run a simulation without pulsar wind nor Be wind, taking into account only the gravitational effect of the pulsar on the Be disk. In this simulation, the gas particles are ejected at a constant rate from the equatorial surface of the Be star, which is tilted in a direction consistent with multi-waveband observations. We run the simulation until the Be disk is fully developed and starts to repeat a regular tidal interaction with the pulsar. Then, we turn on the pulsar wind and the Be wind. We run two simulations with different wind mass-loss rates for the Be star, one for a B2V type and the other for a significantly earlier spectral type. Although the global shape of the interaction surface between the pulsar wind and the Be wind agrees with the analytical solution, the effect of the pulsar wind on the Be disk is profound. The pulsar wind strips off an outer part of the Be disk, truncating the disk at a radius significantly smaller than the pulsar orbit. Our results, therefore, rule out the idea that the pulsar passes through the Be disk around periastron, which has been assumed in the previous studies. It also turns out that the location of the contact discontinuity can be significantly different between phases when the pulsar wind directly hits the Be disk and those when the pulsar wind collides with the Be wind. It is thus important to adequately take into account the circumstellar environment of the Be star, in order to construct a satisfactory model for this prototypical TeV binary.
\end{abstract}

\section{Introduction}
\label{sec:intro}

PSR B1259$-$63/LS~2883 is a massive binary system 
consisting of a 48-ms radio pulsar and a Be star 
with a circumstellar disk \citep{john92}. 
The system is one of only three 
binaries 
from which periodic TeV gamma-ray emission has been detected
(the others are LS~I~+61$-$303 and LS~5039; e.g., \cite{par08}).
Among these TeV gamma-ray binaries, PSR~B1259$-$63/LS~2883 has 
several distinctive properties.
First, the compact star of this system is identified 
as a radio pulsar, namely neutron star, while the nature of the compact object of the other
two sources is still under debate.
Second, the orbital period (3.4\,yr)
 is very long compared to the other systems (3.9\,d for LS~5039 and 26.5\,d for LS~I~+61$~$303), 
 and the eccentricity is so high that
the binary separation at periastron is less than a tenth of  
that at apastron \citep{john94}.
 Third, nonthermal emission is detected in X-ray and radio wave bands 
 as well, and light curves of all bands including TeV gamma-rays
 exhibit multiple-peaked features, 
 in most cases one before periastron and another after it,
 although the curves differ orbit to orbit
 \citep{chernyakova09, aha09}.

Considering the passage of 
 the neutron star through a misaligned Be disk,
 shock acceleration in the disk-pulsar interaction region has been proposed
as the mechanism for double-peaked light curves.
  Modeling of gamma-ray emission was first discussed 
 by \citet{tav94} for MeV energies following the CGRO observation.
 They considered shock acceleration in the interaction region, 
 and showed that the double-peaked feature appeared 
in X-ray and MeV gamma-ray ranges.
 Much theoretical work for high-energy emissions  followed 
(e.g. \cite{tav97}; \cite{bal00}; \cite{mur03}) and 
the emission in TeV energy range has been discussed in \citet{bal00} 
in particular. 
 Assuming a disk-like outflow around the Be star, \citet{kaw04} 
 proposed a double-peaked light curve in TeV range,
 which was followed by the TeV gamma-ray detection  
 with a marginal pre-periastron peak and a clear post-periastron peak
 \citep{aha05}. As shown in the radio observations, 
 the light curve varies from orbit to orbit and 
 the TeV double-peak was not clearly seen in the next 2007 periastron 
 passage. \citet{aha09} suggests that  
 the detection of the TeV emission 50 days prior to the periastron 
 disfavors that the pulsar-disk interaction is 
 the primary TeV emission mechanism. 

  It can be pointed out that  all the previous theoretical studies 
  have either neglected the recent progress in Be star research 
 and adopted an out-of-dated view of the Be disk 
 or neglected even the presence of the Be disk. 
 Given that the gas pressure in the Be disk is much
 higher than the ram pressure of the Be wind, it is  important to take
 into account the interaction between the pulsar wind and the Be disk
 adequately. In this series of papers, 
  we will study the high energy emission from
 PSR B1259$-$63/LS~2883, for the first time based on 3-D numerical
simulations with the latest model of the Be disk.
  As a first step, this paper explores the hydrodynamic interaction
between the pulsar wind and the circumstellar environment of the Be star.

The structure of the paper is as follows. In section~2, we first describe
the Be disk model and the numerical method. We then discuss the viscous evolution
of the Be disk under the tidal interaction with the pulsar,
based on simulations where only the tidal interaction is taken into account.
In section~3, we show the results from simulations that also take account of
both the pulsar wind and the Be wind, and examine the effects of the pulsar wind on 
the circumstellar environment of the Be star.
In section~4, we discuss implications for multi-wavelength observations 
and summarize the conclusions.

\section{Tidal Effect of the Pulsar on the Be Disk}
\label{sec:sim-tidal}

\subsection{Viscous decretion disk model for Be stars}
\label{disk-model}
In this subsection, we briefly summarize key observational features of
Be stars and describe the up-to-date model of the Be disk that is
widely accepted in the Be star community (see \cite{por03} for more
detailed descriptions of Be stars and their circumstellar disks).

A Be star has a two-component extended atmosphere, a polar wind region
and an equatorial disk region. The former consists of 
a low-density fast flow ($\sim 10^3\,{\rm km}\,{\rm s}^{-1}$) emitting UV
radiation. The wind structure is well explained by the line-driven
wind model (\cite{cak75}; \cite{fri86}).  On the other hand, the
equatorial disk consists of a high-density plasma, from which the
optical emission lines and the IR excess arise. The relationship
between the disk size resolved with the optical interferometers
\citep{qui94,qui97} and the separation of the two peaks of the
H$\alpha$ line profile is in agreement with that expected for
a Keplerian disk. The outflow in the Be disk is very subsonic. 
\authorcite{han94} (\yearcite{han94}, \yearcite{han00}) 
and \citet{wat94} showed that the radial
velocity of the disk is smaller than a few ${\rm km\,s}^{-1}$, at
least within $\sim 10$ stellar radii. Thus, it is unlikely that the Be
disk is formed by channeling/focusing of the polar wind. Indeed, it is
established that the Be disk cannot be modeled as a low velocity wind.

One model which can reproduce all key features of the Be disk and is
thus widely accepted, is the viscous decretion disk model proposed by
\citet{lee91} and developed by \citet{por99} and others 
(\cite{oka01}, \yearcite{oka07}; \cite{car06}; \cite{car10}). The model assumes that the
star can eject material with the Keplerian velocity at the stellar
equatorial surface. The ejected material then drifts outward by 
viscous diffusion and forms a geometrically thin Keplerian disk. Recent
Monte Carlo rediative transfer simulations show that such a disk has a temperature
approximately constant at $\sim 60\%$ of the stellar effective
temperature, except in an inner optically-thick region where the disk
is significantly cooler \citep{car06}. It should be noted that the
viscous decretion disk model also provides a firm basis for the study of
tidal interaction between the Be disk and the neutron star in Be/X-ray
binaries \citep{no01, on01}.

\subsection{Numerical model}
\label{num-model}
The simulation presented below was performed with a three dimensional 
Smoothed Particle Hydrodynamics (SPH) code. The code is basically the
same as that used by \citet{oka02} (see also \cite{bat95}). Using a
variable smoothing length, the SPH equations with a standard
cubic-spline kernel are integrated with an individual time step for
each particle. In our code, the Be disk is modeled by an ensemble of
gas particles with negligible self-gravity. For simplicity, the gas
particles are assumed to be isothermal at $0.6\, T_{\rm eff}$ with $T_{\rm
 eff}$ being the effective temperature of the Be star. On the other
hand, the Be star and the neutron star are represented by sink
particles with the appropriate gravitational mass. Gas particles that
fall within a specified accretion radius are accreted by the sink
particle.

In simulations shown in this section, 
the numerical viscosity is adjusted so as to keep the Shakura-Sunyaev
viscosity parameter $\alpha_{\rm SS}=0.1$, using the approximate relation
$\alpha_{\rm SS} = 0.1\, \alpha_{\rm SPH}\, h/H$ and $\beta_{\rm SPH}=0$
\citep{oka02}, where $\alpha_{\rm SPH}$ and $\beta_{\rm SPH}$ are 
the artificial viscosity parameters
\footnote{The artificial viscosity commonly used in SPH consists of two terms:
a term that is linear in the velocity differences between particles, 
which produces a shear and bulk viscosity, and a term that is quadratic 
in the velocity differences, which is needed to eliminate particle interpenetration 
in high Mach number shocks. The parameters $\alpha_{\rm SPH}$ and $\beta_{\rm SPH}$
control the linear and quadratic terms, respectively.},
and $h$ and $H$ are the smoothing length of individual particles 
and the scale-height of the Be decretion disk, respectively.
No effect of pulsar wind or Be wind is
taken into account in these simulations.

\begin{figure}[!t]
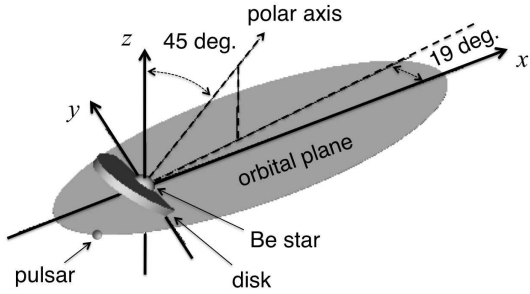

  \centerline{\FigureFile(0.8\hsize,0.7\hsize){figure1.eps}}
    \caption{The model configuration of the VHE gamma-ray binary PSR B1259$-$63/LS~2883.
    The binary orbit is in the $x$-$y$ plane. The polar axis of Be star is tilted 
    by $45\arcdeg$ to the $z$-axis and has $19\arcdeg$ of azimuth of tilt to
    the $x$-axis.}
  \label{fig:config}
\end{figure}

We set the binary orbit in the $x$-$y$ plane with the major axis along
the $x$-axis. At $t=0$, the pulsar is at the apastron (Phase $=$ 0.5). 
It orbits about the Be star with the orbital period
$P_\mathrm{orb}=1236.79\,\mathrm{d}$ and the eccentricity $e=0.87$. It
has long been expected that the Be disk is inclined from the orbital
plane, because the pulsed radio emission disappears for about five weeks centered on
the epoch of periastron, e.g.,
$\sim 18$\,d prior to the 2004 periastron through $\sim 16$\,d after it 
\citep{john05}.
We set the tilt angle between the binary orbital axis and
the Be star's polar axis to $45\arcdeg$  as a parameter
 and the azimuth of tilt, i.e., the azimuthal angle of the Be star's polar axis
from the $x$-axis (the direction of apastron) to $19\arcdeg$ 
as suggested in \citet{chernyakova06} (see Fig.~\ref{fig:config}). 
In order to emulate the mass ejection from a Be star,
we inject gas particles just outside the stellar
equatorial surface at the rate of $3.5 \times
10^{-9}M_{\odot}\,\mathrm{yr}^{-1}$, which gives rise to a typical 
disk base density of $10^{-11}\,{\rm g~cm}^{-3}$. 
Once injected, gas particles
interact with each other. As a result, most injected particles
fall onto the Be star by losing angular momentum, but a small
fraction of particles drift outwards, getting the angular momentum
from the other particles, and form a disk.

Table~\ref{tbl:parameters} summarizes the parameters adopted in the following
simulations. Note that not all parameters in Table~\ref{tbl:parameters} 
are independent. For instance, all Be disk parameters except for the 
tilt angles are derived from other parameters. 

\begin{table}
\caption{Model parameters}
 \begin{tabular}{p{0.45\textwidth}}
 \hline
 Be star parameters\\
 \hline
 Mass of the Be star $M_{*} =$ $10\,M_\odot$~$^{\rm a}$ \\
 Radius of the Be star $R_{*} =$ $6\,R_\odot$~$^{\rm a}$  \\
 Effective temperature $T_{\rm eff} = 22,800\,{\rm K}$~$^{\rm b}$\\
 Critical velocity $V_\mathrm{crit}(R_*) = 563.9\,{\rm km\,s}^{-1}$ \\
 Wind velocity $V_{\rm wind} = 10^{3}\,{\rm km~s}^{-1}$\\
 Wind mass-loss rate
 \begin{displaymath}
 \hspace*{3.5em}\dot{M}_{\rm wind} = 
               \left\{ \begin{array}{ll}
                 10^{-9}M_{\odot}{\rm yr}^{-1}
                      & {\rm (\lq\lq weak'' \; case)}\\
                 10^{-8}M_{\odot}{\rm yr}^{-1}
                      & {\rm (\lq\lq strong'' \; case)}\\
              \end{array} \right.
 \end{displaymath}
Mass injection rate to the disk
 \begin{displaymath}
 \hspace*{3.5em}\dot{M}_{\rm disk} = 3.5 \times 10^{-9}M_{\odot}{\rm yr}^{-1}\,^{\rm c}
 \end{displaymath}\\
 \hline
 Be disk parameters\\
 \hline
 Disk temperature $T_{\rm d} = 13,680\,{\rm K}$~$^{\rm d}$  \\
 Disk thickness $H(R_*)/R_* = 0.024$ \\
 Sound speed $c_{\rm s} = 13.8\,{\rm km\,s}^{-1}$ \\
 Initial tilt angles: Misalignment by $45\arcdeg$ from the binary orbital axis,
   with the polar axis projection
   of $19\arcdeg$ from apastron~$^{\rm e}$ 
   (see Fig.~\ref{fig:config}) \\
 \hline
 Pulsar and orbital parameters\\
 \hline
 Pulsar wind power $\dot{E}_\mathrm{PSR} = 8.2 \times 10^{35}\,\mathrm{erg\,s}^{-1}$~$^{\rm a}$ \\
 Orbital period $P_{\rm orb} = 1236.79\,{\rm d}$ ($= 3.386\,{\rm yr}$)~$^{\rm f}$   \\
 Orbital eccentricity $e = 0.87$~$^{\rm a}$ \\
 Mass ratio $q = 0.14$ ($M_\mathrm{PSR} = 1.4\,M\odot$)~$^{\rm a}$  \\
 Semimajor axis $a = 181.82\,R_*$
       $= 7.59\cdot 10^{13}\,{\rm cm}$~$^{\rm a}$ \\
 \hline
$^{\rm a}$ \citet{john94}\\
$^{\rm b}$ The effective temperature of a B2V star \citep{cox00}\\ 
$^{\rm c}$ Mass ejection rate that gives the typical disk base density of 
$10^{-11}\,{\rm g~cm}^{3}$\\
$^{\rm d}$ Assumed to be isothermal at $0.6\,T_{\rm eff}$ \citep{car06}.\\
$^{\rm e}$ \citet{chernyakova06}\\
$^{\rm f}$ Taken from \cite{john94}. Recent timing analyses give $P_{\rm orb} = 1236.72\,{\rm d}$ \citep{wan04}.\\
\end{tabular}
\label{tbl:parameters}
\end{table}

\subsection{Long-term evolution of the Be disk under no influence of the pulsar wind}
\label{sec:disk-evol}

\begin{figure}
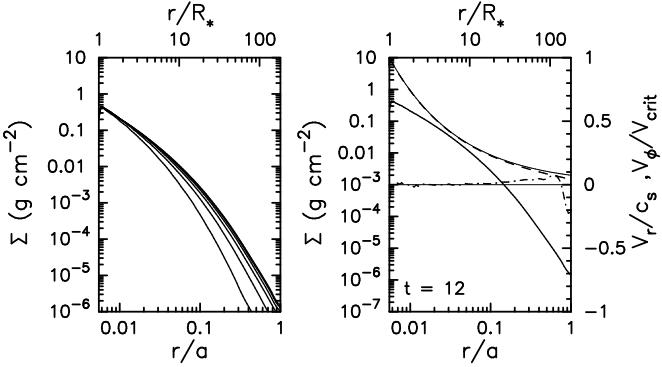

  \FigureFile(\hsize,\hsize){figure2.eps}
    \caption{Left: Surface density evolution from $t = $0 to 12
    in units of $P_{\rm orb}$ in a simulation where 
    only the tidal interaction is taken into account.
    The interval of time between adjacent contours
    is 2 ($t =$ 2, 4, $\ldots$, 12 from bottom).
    Right: Disk structure at $t = 12$.
    The thick solid, dashed, and dash-dotted lines denote
    the surface density, the azimuthal velocity
    normalized by the stellar critical velocity, and the radial Mach number,
    respectively. For comparison, the Keplerian rotation distribution
    is shown by the thin solid line.
    In both panels, the
    density is integrated vertically and averaged azimuthally, while
     the velocity components are averaged
     vertically and azimuthally.}
    \label{fig:rdist}
\end{figure}

Figure~\ref{fig:rdist} shows the evolution of surface density from $t
= 0$ to $12\, P_{\rm orb}$ (panel a) and the disk structure at $t=12\,
P_{\rm orb}$ (panel b). In Fig.~\ref{fig:rdist}, the volume density is
integrated vertically and averaged azimuthally, while the velocity
components are averaged vertically and
azimuthally. Figure~\ref{fig:rdist}(a) exhibits how a decretion disk
forms if there is no influence of the pulsar wind. 
The disk density gradually increases with time and approaches
an asymptotic distribution. The timescale of the disk growth is the
viscous timescale, which is given by $[\alpha_\mathrm{SS} (H/r)^{2}
\Omega_\mathrm{K}]^{-1}$, where
$\Omega_\mathrm{K}=(GM_{*}/R_{*}^{3})^{1/2}$ is the Keplerian angular
frequency. For parameters summarized in Table~\ref{tbl:parameters},
this timescale is  $\sim
12(\alpha/0.1)^{-1}(r/10R_{*})^{1/2}\,\mathrm{yr}$. Thus, at $t \sim
12\,P_\mathrm{orb} \sim 40\,\mathrm{yr}$, the disk is already fully
developed, as seen in Fig.~\ref{fig:rdist}(a).

We also note from Fig.~\ref{fig:rdist}(a) that the disk extends beyond
the periastron separation ($r \sim 24R_{*}$) with no break in the
radial density distribution. This suggests that the Be disk in long-period,
highly-eccentric systems like PSR B1259$-$63 is not truncated solely
by the tidal torques of the neutron star, 
unlike disks in other Be-neutron star binaries with shorter orbital periods
and lower eccentricities. 
In such systems, the neutron star passes through, and thus strongly
disturbs, the Be disk around every periastron passage, as we will see in
section~\ref{sec:tidal-result}.

As discussed in section~\ref{disk-model}, the Be disk is Keplerian. 
Fig.~\ref{fig:rdist}(b) shows that the
angular velocity (the dashed line) for $r \lesssim 100R_{*}$ is
indistinguishable from the Keplerian rotation (the thin solid line).
The radial velocity, $V_r$, is very subsonic. It increases with
radius, but even at $~100R_{*}$, $V_r$ is still of the order of
$\sim 0.1c_\mathrm{s} \sim 1\,\mathrm{km\,s}^{-1}$,
where $c_\mathrm{s}$ is the isothermal sound speed in the Be disk.
These are typical features of viscous decretion disks
around Be stars (e.g., \cite{lee91}).

\subsection{Tidal interaction around periastron passage}
\label{sec:tidal-result}

\begin{figure*}[th!]
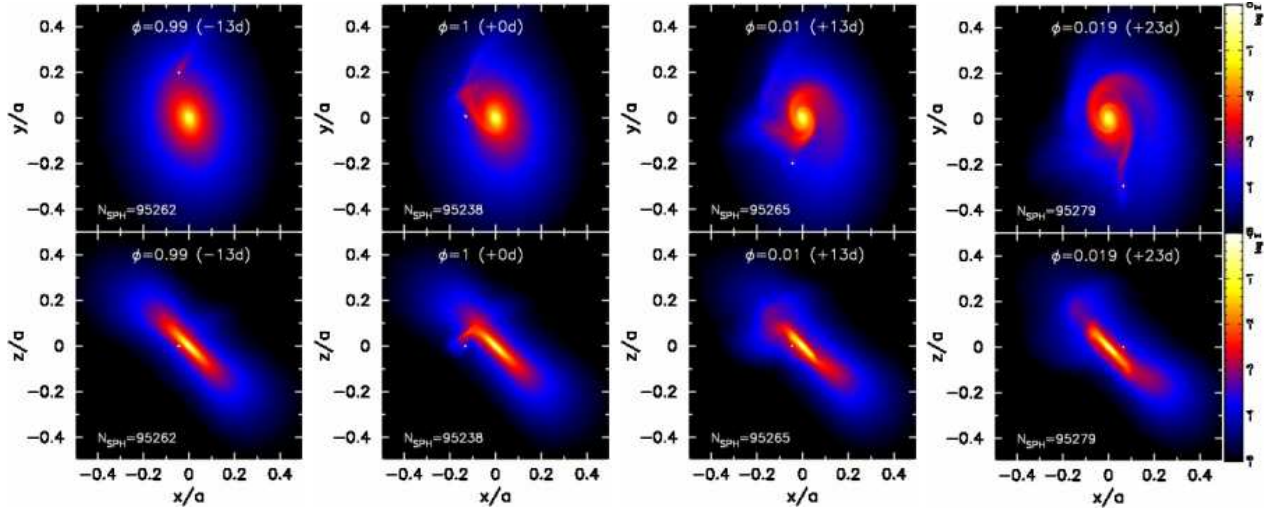

  \begin{center}
  \FigureFile(0.22\hsize,0.22\hsize){figure3-1.eps}
  \FigureFile(0.22\hsize,0.22\hsize){figure3-2.eps}
  \FigureFile(0.22\hsize,0.22\hsize){figure3-3.eps}
  \FigureFile(0.249\hsize,0.22\hsize){figure3-4.eps}
  \end{center}
    \caption{Snapshots over a month around periastron passage
    for $\rho_{0}=10^{-11}\mathrm{g\,cm}^{-3}$, where
    $\rho_{0}$ is the base density of the Be disk.
    The upper panels show the column density along the $z$-axis 
    (binary orbital axis) in cgs units,
    while the lower panels the column density along the $y$-axis 
    (minor axis of the binary orbit). 
    Annotated in each panel are the
    phase and time measured from periastron passage and the number of
    SPH particles.}
  \label{fig:snapshots1}
\end{figure*}

Figure~\ref{fig:snapshots1} provides snapshots over a month from
13\,d before periastron passage ($\phi=0.99$) to 23\,d after it
($\phi=0.02$). The upper and lower panels show respectively the column density 
in cgs units along the binary orbital axis ($z$-axis) and 
the minor axis of the binary orbit ($y$-axis).
The first disk-passing event of the pulsar starts about 
20 days prior to periastron passage,
while the second event starts $\sim 12$ days after it. 
Each event lasts for $\sim 2$ weeks. 
As seen in the figure, the Be disk is strongly disturbed by these
events. The disk is temporarily warped at each of these events
via the tidal interaction with the pulsar. At the same time, the tidal
interaction around periastron excites a two-armed spiral wave, via
which angular momentum is removed from the disk (see, e.g.,
\cite{art94}). This angular momentum transfer causes shrinking of
the Be disk. Its radius gradually recovers afterward by viscous
diffusion \citep{oka02}. 
Note that because of the cumulative effect of the tidal torques,
the disturbance in the Be disk is larger after periastron passage,
despite that the distance between the Be disk and pulsar is closest at the first
disk-passing event.

\section{Effects of the Pulsar Wind on the Circumstellar Environment of the Be star}
\label{sim-wind}

\subsection{Numerical setup}
In order to see how the pulsar wind interacts with the Be disk and wind,
we carried out simulations where both the pulsar and Be winds are now taken into account, 
using the data from the above tidal simulation. In these new wind simulations,
we turned on the pulsar and Be winds at a certain time, after the Be disk was fully developed 
in the tidal simulation. We started the wind simulation at $t=11.44\,P_{\rm orb}$, 
74 days prior to periastron, much longer than the crossing time of the Be wind 
over the simulation volume, $\sim 10\,{\rm d}$. 
The relativistic pulsar wind crosses the simulation volume much faster. 
Although in the following we model the pulsar wind by a flow 
with the non-relativistic speed of $10^{4}\,{\rm km s}^{-1}$, 
the crossing time for such a \emph{slow} wind is only $\sim 1\,{\rm d}$. 
Therefore, the starting time does not matter as long as it is at least several tens of days 
before periastron.

In the tidal simulation, we emulated the Shakura-Sunyaev viscosity, adopting variable $\alpha_{\rm SPH}$ over space and time and fixing $\beta_{\rm SPH}$ to 0.
In the wind simulation, however, this method would fail, allowing the particle interpenetration at strong shocks. Therefore, we adopted the standard values of
the artificial viscosity parameters in the wind simulation, i.e., $\alpha_{\rm SPH}=1$ and 
$\beta_{\rm SPH}=2$. 
In the wind simulations, the energy equation is also different from the tidal simulation. 
In the latter, the gas was assumed to be isothermal, in order to emulate the temperature distribution in the Be disk. However, the purpose of the wind simulation is to study the structure of the wind-wind and wind-disk collisions. Therefore, in the wind simulation,
we take account of optically thin radiative cooling.
Numerical implementation of radiative cooling is done by adopting 
\citet{tow09}'s Exact Integration Scheme for radiative cooling,
with the cooling function generated with CLOUDY 90.01 for an optically thin plasma 
with solar abundances \citep{fer96}.

\begin{figure*}[!t]
  \begin{center}
  (a) $\dot{M}_{\rm wind}=10^{-9} M_\odot\,{\rm yr}^{-1}$\\
  \FigureFile(0.22\hsize,0.22\hsize){figure4a-1.eps}
  \FigureFile(0.22\hsize,0.22\hsize){figure4a-2.eps}
  \FigureFile(0.22\hsize,0.22\hsize){figure4a-3.eps}
  \FigureFile(0.249\hsize,0.22\hsize){figure4a-4.eps}

  \bigskip  
  (b) $\dot{M}_{\rm wind}=10^{-8} M_\odot\,{\rm yr}^{-1}$\\
  \FigureFile(0.22\hsize,0.22\hsize){figure4b-1.eps}
  \FigureFile(0.22\hsize,0.22\hsize){figure4b-2.eps}
  \FigureFile(0.22\hsize,0.22\hsize){figure4b-3.eps}
  \FigureFile(0.249\hsize,0.22\hsize){figure4b-4.eps}
  \end{center}
    \caption{Snapshots over a month around periastron passage
    from simulations in which both the pulsar wind and the Be wind are taken into account:
    (a) $\dot{M}_{\rm wind}=10^{-9} M_\odot\,{\rm yr}^{-1}$ and 
    (b) $\dot{M}_{\rm wind}=10^{-8} M_\odot\,{\rm yr}^{-1}$.
    The mass-loss rate via the Be disk is $3.5 \times 10^{-9} M_\odot\,{\rm yr}^{-1}$,
    where $\dot{M}_{\rm wind}$ is the mass-loss rate through the Be wind.
    In each panel,
    $N_1$, $N_2$, and $N_3$ annotated at the lower left corner are
    the numbers of particles in the Be wind, the pulsar wind, and 
    the Be disk, respectively.}
  \label{fig:snapshots2}
\end{figure*}

With a large binary separation, the winds are likely to collide after the Be wind reaches 
a terminal speed. Thus, for simplicity, we assume that the winds coast 
without any net external forces, assuming in effect that gravitational forces are 
either negligible (i.e. for the pulsar wind) or are canceled by radiative driving terms 
(i.e. for the Be wind).
The relativistic pulsar wind is emulated by a non-relativistic $10^{4}\,{\rm km s}^{-1}$ wind 
with the adjusted mass-loss rate so as to give the same momentum flux as a relativistic flow 
with the assumed energy, as in \citet{rom07}.

In the following, for simplicity, we assume that all the spin down energy 
$\dot{E}_{\rm PSR}=8.2 \times 10^{35}\,{\rm erg~s}^{-1}$ goes to 
the kinetic energy of a spherically symmetric pulsar wind. 
We also assume the Be wind to be spherically symmetric. 
Given a recent high-resolution spectroscopic study of LS~2883 suggesting 
a significantly earlier spectral type than the conventionally used spectral type B2V 
\citep{neg11}, we take two different values of mass-loss rates, 
$10^{-9}\,M_{\odot}\,{\rm yr}^{-1}$ for the conventional spectral type B2V 
and $10^{-8}\,M_{\odot}\,{\rm yr}^{-1}$ for an earlier spectral type. 
The mass injection rate to the Be disk is fixed to $3.5 \times 10^{-9}\,M_{\odot}\,{\rm yr}^{-1}$.

\subsection{Interaction between the pulsar wind and the Be disk and wind}

\begin{figure*}[!th]
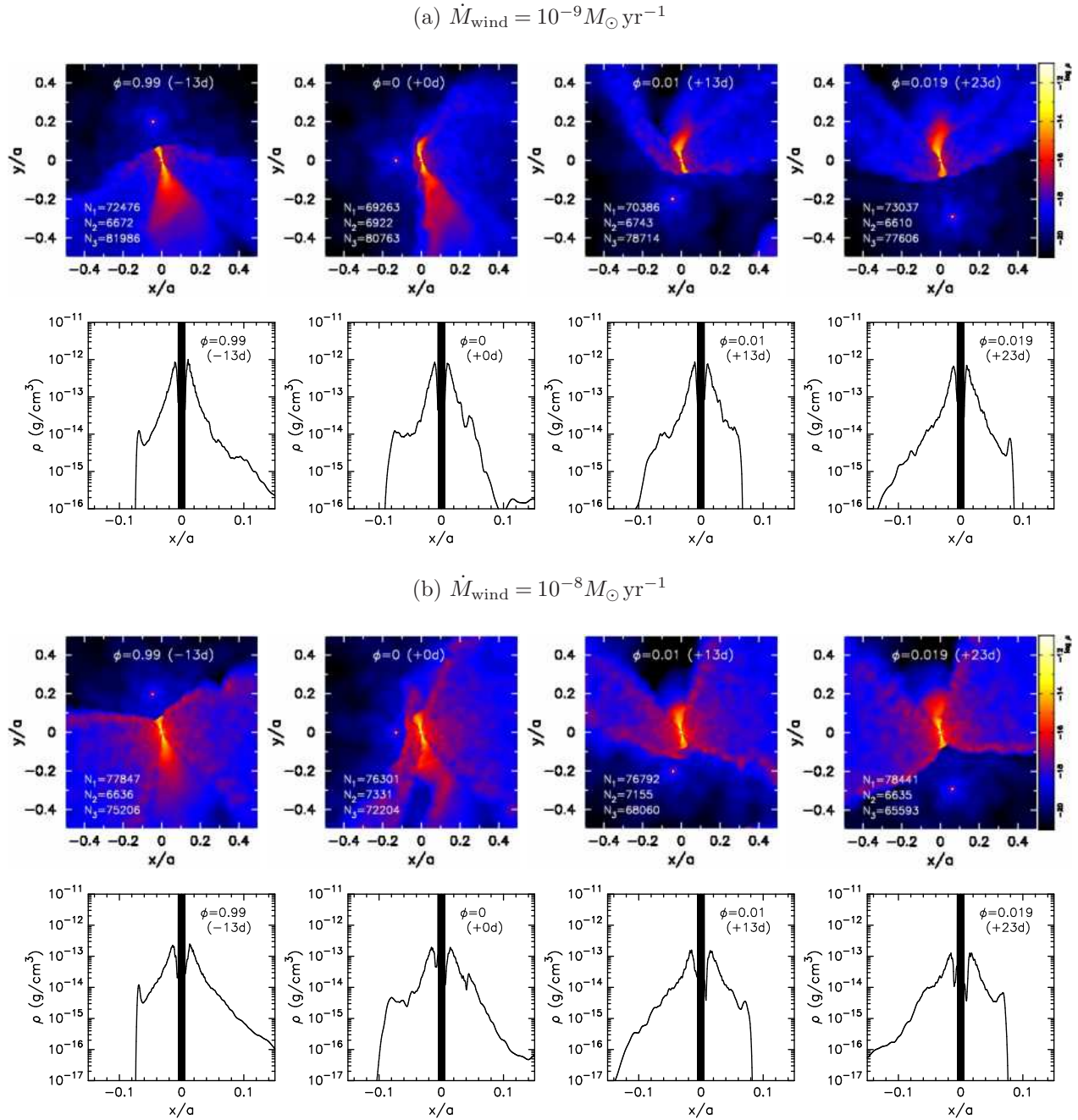

  \begin{center}
  (a) $\dot{M}_{\rm wind}=10^{-9} M_\odot\,{\rm yr}^{-1}$\\
  \FigureFile(0.22\hsize,0.22\hsize){figure5a-1u.eps}
  \FigureFile(0.22\hsize,0.22\hsize){figure5a-2u.eps}
  \FigureFile(0.22\hsize,0.22\hsize){figure5a-3u.eps}
  \FigureFile(0.249\hsize,0.22\hsize){figure5a-4u.eps}
  \FigureFile(0.22\hsize,0.22\hsize){figure5a-1l.eps}
  \FigureFile(0.22\hsize,0.22\hsize){figure5a-2l.eps}
  \FigureFile(0.22\hsize,0.22\hsize){figure5a-3l.eps}
  \FigureFile(0.22\hsize,0.22\hsize){figure5a-4l.eps}

  \bigskip  
  (b) $\dot{M}_{\rm wind}=10^{-8} M_\odot\,{\rm yr}^{-1}$\\
  \FigureFile(0.22\hsize,0.22\hsize){figure5b-1u.eps}
  \FigureFile(0.22\hsize,0.22\hsize){figure5b-2u.eps}
  \FigureFile(0.22\hsize,0.22\hsize){figure5b-3u.eps}
  \FigureFile(0.249\hsize,0.22\hsize){figure5b-4u.eps}
  \FigureFile(0.22\hsize,0.22\hsize){figure5b-1l.eps}
  \FigureFile(0.22\hsize,0.22\hsize){figure5b-2l.eps}
  \FigureFile(0.22\hsize,0.22\hsize){figure5b-3l.eps}
  \FigureFile(0.22\hsize,0.22\hsize){figure5b-4l.eps}
  \end{center}
    \caption{Snapshots of the volume density in the orbital plane (upper panels) and 
    along the disk mid-plane (lower panels): 
    (a) $\dot{M}_{\rm wind}=10^{-9} M_\odot\,{\rm yr}^{-1}$ and 
    (b) $\dot{M}_{\rm wind}=10^{-8} M_\odot\,{\rm yr}^{-1}$.
    In the lower panels of each figure, the $x>0$ part
    corresponds to the $x>0$ disk region (lower-right direction along the disk mid-plane)
    in the upper panels.
    Times of the plots are the same as in Figs.~\ref{fig:snapshots1} and 
    \ref{fig:snapshots2}.}
  \label{fig:rho2d+1d}
\end{figure*}

Once winds are turned on, the faster pulsar wind soon fills up the whole simulation volume
of $r \le a$. Then, the slower Be wind pushes the pulsar wind back to the radii 
where the ram pressures of both winds are balanced. Without the Be disk,
the shape of the interaction surface would be determined by the wind-wind collision.
A key parameter would, then, be the ratio of wind momentum fluxes, given by 
\begin{equation}
   \eta = \frac{\dot{E}_{\rm PSR}}{\dot{M}_{\rm wind} V_{\rm wind} c},
   \label{eq:eta}
\end{equation}
where $V_{\rm wind}$ and $\dot{M}_{\rm wind}$ are the velocity
and mass loss rate of the Be wind, respectively, and  
$\dot{E}_{\rm PSR}$ is the power of the pulsar wind.
Taking $\dot{E}_{\rm PSR}=8.2 \times 10^{35}\,{\rm erg~s}^{-1}$ for the spherically symmetric pulsar wind and $V_{\rm wind}=10^{3}\,{\rm km\,s}^{-1}$ as the velocity of the Be wind, 
we have $\eta \sim 4.3$ for the \lq\lq weak'' wind case
($\dot{M}_{\rm wind}=10^{-9} M_{\odot}\,{\rm yr}^{-1}$) and 
$\eta \sim 0.43$ for the \lq\lq strong'' wind case 
($\dot{M}_{\rm wind}=10^{-8} M_{\odot}\,{\rm yr}^{-1}$).
Note that the pulsar wind dominates the Be wind in the former case, while the Be wind is about
twice stronger than the pulsar wind in the latter case. Thus, having an accurate spectral type
of the Be star is important to construct any satisfactory model for this VHE gamma-ray binary.

In a simple 2D model that ignores orbital motion, 
the ram pressure balance then implies that, for a binary
separation $D$, the interface should be located at a distance
\begin{equation}
   d = \frac{D}{1+\sqrt{\eta}}
     \sim \left\{ \begin{array}{ll}
               0.60 D & {\rm (strong~wind)}\\
               0.33 D & {\rm (weak~wind)}
               \end{array} \right.
   \label{eq:location}
\end{equation}
from the Be star \citep{SBP92, CRW96}. 
In adiabatic shocks, the interaction surface approaches a cone 
with the half opening angle,
\begin{equation}
    \theta = 2 \left( \tan^{-1} \eta \right)^{1/4}
     \sim \left\{ \begin{array}{ll}
               102\arcdeg & {\rm (strong~wind)}\\
               69\arcdeg & {\rm (weak~wind)}
               \end{array} \right.
    \label{eq:opang}
\end{equation}
\citep{gay09}, measured from the Be star. Thus, the shock is wrapped around the Be star in the weaker wind
case, while it is around the pulsar in the stronger wind case.

In PSR B1259$-$63/LS~2883, the presence of the Be disk adds extra complications
in the interaction with the pulsar wind. Figure~\ref{fig:snapshots2} shows how the interaction
occurs around periastron in the weak (upper figure) and strong (lower figure) wind simulations. 
In each figure, the upper and lower panels respectively present 
the column density along the $z$-axis (binary orbital axis) 
along the $y$-axis (minor axis or the binary orbit).
The change in the Be disk structure from that in the tidal simulation (Fig.~\ref{fig:snapshots1}) is drastic. The pulsar wind now strips off an outer part of the Be disk
on the side facing the pulsar, truncating it at a radius where the gas pressure of 
the disk is roughly comparable to the ram pressure of the pulsar wind.
Since the pulsar wind has a velocity component tangential to the line 
connecting the Be star and the pulsar,
the material in the disk outside this truncation radius is pushed sideways, 
forming a warped filament in the prograde direction.
This feature is more remarkable in the weak wind simulation than in the strong wind simulation, 
because, unlike in the latter simulation, the Be wind in the former simulation 
is too weak to shield the disk from the pulsar wind.

The interaction between the pulsar wind and the Be-star circumstellar environment is 
more clearly seen in Fig.~\ref{fig:rho2d+1d}, in particular in the upper panels,
where the volume density in the binary orbital plane (upper panels) and
along the disk mid-plane (lower panels) is shown at the same phases as in 
Figs.~\ref{fig:snapshots1} and \ref{fig:snapshots2}. 
Comparing Fig.\ref{fig:rho2d+1d}(a) for the weak wind case with 
Fig.\ref{fig:rho2d+1d}(b) for the strong wind case, we note that the effect of the pulsar wind
on the Be disk depends on the relative strength of the Be wind.
In the case of weak Be wind, the Be disk is not only truncated 
but also strongly deformed by the pulsar's ram pressure around periastron. 
In contrast, little deformation is seen in the case of strong Be wind, where
the Be wind shields the Be disk from the pulsar wind.

In both simulations, the truncation of the Be disk is so efficient that 
the disk has a sharp density drop at the outer radius.
At phases of closest encounter of the pulsar with the disk outer radius, 
the density at the outer radius is enhanced by the shock by a factor of several,
as seen in the lower panels of Fig.~\ref{fig:rho2d+1d}.
This sharp truncation of the Be disk is likely a characteristic to be seen only in systems
consisting of a Be star and an object with a strong wind.
Although Be disks in binaries are, in general, tidally truncated, 
the density decrease beyond the disk outer radius is much more gradual in 
the tidal truncation. The density drop near the central star is due to 
ablation by the Be wind.

Despite these features related to the Be disk, 
the location and global shape of the interaction
surface between the pulsar wind and the circumstellar environment of the Be star,
i.e., the Be disk and wind, is consistent with that expected in the above analysis 
using a simple 2D model without the Be disk.
We can see that in Fig.~\ref{fig:rho2d+1d}, where the density 
on the binary orbital plane is shown at the same phases as in 
Figs.~\ref{fig:snapshots1} and \ref{fig:snapshots2}, the apex location and 
conical shape of the interaction surface in the simulation, in general, agree
with those given by equations~(\ref{eq:location}) and (\ref{eq:opang}).

\section{Summary and Discussion}
\label{discussion}

In this paper, we have numerically investigated the hydrodynamic interaction of 
the circumstellar environment of the Be star with the pulsar
in the TeV binary PSR B1259$-$63/LS~2883.
We have broken the computation into two separate, but linked parts. 
The first part considers only the gravitational interaction with the pulsar,
ignoring the effects of both the pulsar and Be star winds. 
One purpose of this part of simulation is to see
how the Be disk evolves by the effect of viscosity and modulates by 
the tidal interaction with the pulsar.
The other purpose is to set up the initial configuration 
for the second part of simulation, which takes account of both pulsar and Be winds.

From the first part, we confirmed the viscous decretion disk scenario that 
the material ejected from the stellar equatorial surface drifts outward by the effect of viscosity, and forms a Keplerian disk. Note that the radial velocity caused by 
viscous diffusion is very subsonic,
so it takes more than a decade for this very wide system to have 
a fully developed Be disk.
Without the pulsar wind, the Be disk is not truncated in such a highly eccentric system, unlike tidally truncated Be disks in binaries with lower eccentricities. Then, the pulsar passes through the disk twice an orbit, causing the tidal stream between the disk and the pulsar.

The second part focuses on the interaction between the pulsar wind and the circumstellar material of the Be star around periastron. 
It shows that once winds from the pulsar and the Be star
are switched on, their effect soon becomes apparent in the structure of the Be disk.
The pulsar truncates the Be disk on the side facing the pulsar, sweeping up an outer part to a dense filament. The resulting structure of the disk is strongly asymmetric
and phase dependent. The size of the truncated disk is so small that 
the pulsar never passes through the disk.
Despite these features, 
the location and global shape of the interaction
surface between the pulsar wind and the Be wind is well described by a simple 2D model.

We performed numerical simulations of Newtonian hydrodynamics without magnetic fields,
although both of relativity and magnetic fields can play important roles 
on the dynamics of the system. 
A highly relativistic pulsar wind was emulated by a Newtonian wind by equating its momentum flux (thrust). 
Some studies of special relativistic hydrodynamic simulations show 
that highly supersonic flows display extended cocoons with high pressure \citep{marti97}, which
correspond to the shocked pulsar winds in this study. 
The cocoons (or pulsar winds) have smoother local
structure in relativistic simulations than in non-relativistic
simulations, because the inertia of a relativistic flow is smaller than that of a non-relativistic flow with the same thrust \citep{rosen99}.
The global structure of the shocked pulsar wind, however,
will not considerably be changed, since we have tuned the wind momentum flux
in the simulation identical to that of relativistic wind of PSR B1259-63.
The overall structure of the shock is basically determined by the
balance of momentum fluxes of the flows. It is unlikely as well that the magnetic fields strongly
affects the overall interaction feature, given that the kinetic energy is, in general, much greater 
than the magnetic field energy in pulsar winds near termination shocks (e.g. \cite{kennel84}). 
Introduction of magnetic fields into the simulation, however, may be needed to study
the local structure in detail \citep{mckinney06, komissarov09, nagataki09, nagataki10}. 
Therefore, it will be important to extend our code to a relativistic regime (e.g.
\cite{monaghan01, ryu06, rosswog10}) with magnetic fields (e.g. \cite{price05})
in the future so that we can self-consistently study these effects.

It is frequently discussed that the particle acceleration happens at
the shock in the pulsar wind side and the accelerated, non-thermal
electron-positron pairs emit radio to X-ray photons through synchrotron
radiation while VHE (GeV-TeV) gamma-rays are produced by inverse-Compton
scatterings of soft photons from the Be star \citep{tav97,
bednarek08, takata09, naito10}. It is also claimed that proton acceleration 
at the shock in the Be star side and VHE gamma-rays from pion decays that
are produced through proton-proton interactions may be relevant to 
explain VHE gamma-rays \citep{kaw04,chernyakova06}. From the point of view
of the time-correlation among radio, X-rays, and VHE gamma-rays
\citep{chernyakova06}, hadronic scenario may be favored. This is
because the anti-correlation of fluxes between synchrotron emission and
inverse-Compton scatterings is expected for the leptonic scenario, which
seems to contradict the observations \citep{tav97}. To explain the observed 
correlation between X-ray and VHE emissions, therefore, 
\citet{takata09} invoked the scenario
that the observed emission in different orbital phases emanate from different magnetic field lines,
giving rise to the pulsar wind parameters changing with the orbital phase or 
that the power law index of the accelerated particles varies with the orbital phase. 
However, since these discussions depend 
on one-zone models adopted, the anti-correlation may not appear even for the
leptonic scenario when we investigate a more realistic situation using
the results from our simulations. For example, the previous 
leptonic models have  assumed  that the pulsar is confined by 
the  shock \citep{tav97, takata09},
 whereas the result from the present simulations suggests that 
the shock geometry is sensitive to the mass loss rate from the Be star and 
to the orbital phase, as Figure~5 shows. It is thus essential to calculate 
the high energy emission processes with a more realistic situation.
We are planning to investigate it as
our next step.

While a pulsar wind $+$ stellar wind interaction model has also been proposed for other high-mass binaries with GeV-TeV gamma-ray emission (e.g., LS~I +61 303 and LS~5039), PSR B1259$-$63 is so far the only such system with direct detection of pulsed radio emission. \citet{john96} showed that this radio emission disappears for a period of about 5 weeks centered on the periastron epoch of 1994 January 9. They attributed this as \lq\lq most likely due to a combination of free-free absorption and severe pulse scattering in the Be-star disc''. But note that GHz radio emission is expected also to undergo strong free-free absorption by the much lower-density Be-star {\em wind}, since for typical wind parameters, the associated free-free optical depth can be of several {\em thousand} at orbital distances around an AU [see, e.g., equation~(4) of \citet{tor10}]. As such, the detection of radio pulses in PSR B1259$-$63 may actually be due to an observer perspective that is nearly aligned to the major axis of the binary orbit, such that during most of the orbit around apastron, the observed radio emission propagates mostly through the low-density shock cone surrounding the relativistic pair wind. In this scenario, the disappearance of radio emission near periastron could actually be attributed to the rapid rotation of the shock cone away from the viewer's line of sight, causing strong free-free attenuation of the radio emission by the Be-star wind and its associated shock cone. A similar viewer perspective  has been used to explain the several- week-long attenuation of X-ray emission in the colliding wind binary $\eta$ Carinae \citep{oka08}. Our future work will further explore this shock-cone scenario as an alternative to the standard Be-disk attenuation model for the periastron disappearance of pulsar radio emission.

Observations of TeV binaries in high energy bands provide important information on 
the type of interaction and the emission mechanisms. For TeV binaries with Be stars, however, 
the high energy bands are not the sole window to probe these systems.
As mentioned in section~2, the optical emission lines and infrared excess arise from the Be disk, 
while the Be wind emits UV radiation.
Observations in these bands are thus suitable to study the effects of the pulsar wind 
on the circumstellar environment of the Be star. 
Particularly, spectroscopic and polarimetric observations in the optical and infrared 
can provide valuable information on the structure and dynamics of the disturbed Be disk,
which gives clues for the interaction between the pulsar wind and the Be disk.
UV observations, on the other hand, are adequate for studying 
the strongly asymmetric Be wind around periastron (see Fig.~\ref{fig:rho2d+1d}
for the asymmetric structure of the Be wind).
For better understanding of TeV binaries with Be stars, including PSR B1259$-$63,
multi-band observations are highly desirable.

\bigskip

We thank Ignacio Negueruela and Mark Rib\'{o} for letting us know
about their newly determined spectral type of LS~2883.
The SPH simulations were performed on HITACHI SR11000 at the Information 
Initiative Center (iiC), Hokkaido university, Sapporo, Japan. This work
was partially supported by the iiC collaborative research program 2009-2010,
the Grant-in-Aid for the Global COE Program "The Next Generation of
Physics, Spun from Universality and Emergence" from the Ministry of
Education, Culture, Sports, Science and Technology (MEXT) of Japan,
and the Grant-in-Aid for Scientific Research 
(18104003, 19047004, 19104006, 19740100, 20540236, 21105509, 21540304, 22340045, 22540243).
SPO acknowledges partial support from grant \#NNX11AC40G from NASA's Astrophysics Theory Program.

\label{lastpage}

 \end{document}